\documentclass[12pt]{article}
\usepackage{epsfig}
\def\f{\frac}

\def\nn{\nonumber}
\def\be{\begin{equation}}
\def\ee{\end{equation}}
\newcommand{\bea}{\begin{eqnarray}}
\newcommand{\eea}{\end{eqnarray}}
\newcommand{\bdm}{\begin{displaymath}}
\newcommand{\edm}{\end{displaymath}}

\def\prog{SusyBSG}
\def\theprog{{\tt \prog}}
\def\vernum{{\tt 1.1}}
\long\def\symbolfootnote[#1]#2{\begingroup%
\def\thefootnote{\fnsymbol{footnote}}\footnote[#1]{#2}\endgroup}

\def\hlf{\frac{1}{2}}

\def\drbar{\overline{\rm DR}}
\def\msbar{\overline{\rm MS}}
\def\sq2{\sqrt{2}}

\def\as{\alpha_s}
\def\at{\alpha_t}
\def\ab{\alpha_b}
\def\oas{{\cal O}(\as)}
\def\oat{{\cal O}(\at)}
\def\oab{{\cal O}(\ab)}
\def\Bsg{B\to X_{\!s}\, \gamma}
\def\bsg{b \to s \gamma}
\def\bratio{{\rm BR}[\Bsg]}

\newcommand{\smallz}{{\scriptscriptstyle Z}} 
\newcommand{\smallw}{{\scriptscriptstyle W}} %

\newcommand{\mz}{m_\smallz}
\newcommand{\mw}{m_\smallw}

\newcommand{\muw}{\mu_\smallw}
\newcommand{\mut}{\mu_t}
\newcommand{\susy}{\rm{\scriptscriptstyle {\rm SUSY}}}
\newcommand{\mumfv}{\mu_{{\scriptscriptstyle {\rm MFV}}}}
\newcommand{\mususy}{\mu_{\susy}}
\newcommand{\msusy}{M_{\scriptscriptstyle S}}

\def\mt{m_t}
\def\mb{m_b}

\def\mgl{m_{\tilde{g}}}
\def\mhc{m_{H^\pm}}
\def\mchi{m_{\chi_i}}

\def\vckm{V^{\rm \scriptscriptstyle CKM}}
\def\vckmc{V^{{\rm \scriptscriptstyle CKM}\,*}}
\def\vckmd{V^{{\rm \scriptscriptstyle CKM}\,\dagger}}

\def\msbu{m_{\tilde{b}_1}^2}
\def\msbd{m_{\tilde{b}_2}^2}

\newenvironment{AppendA}
 {
  \setcounter{section}{0}
  \setcounter{equation}{0}
  
 }

\newenvironment{AppendB}
 {
  \setcounter{equation}{0}
  
 }

\newenvironment{AppendC}
 {
  \setcounter{equation}{0}
  
 }

\newenvironment{AppendD}
 {
  \setcounter{equation}{0}
  
 }

\newcounter{bla}
\newenvironment{refnummer}{%
\list{[\arabic{bla}]}%
{\usecounter{bla}%
 \setlength{\itemindent}{0pt}%
 \setlength{\topsep}{0pt}%
 \setlength{\itemsep}{0pt}%
 \setlength{\labelsep}{2pt}%
 \setlength{\listparindent}{0pt}%
 \settowidth{\labelwidth}{[9]}%
 \setlength{\leftmargin}{\labelwidth}%
 \addtolength{\leftmargin}{\labelsep}%
 \setlength{\rightmargin}{0pt}}}
 {\endlist}

\begin{document}
\thispagestyle{empty}
\vspace*{-15mm}

\vspace*{-8mm}
\begin{flushright}
RM3-TH/07-19\\
DFTT-28/2007\\
LAPTH-1225/07\\
CERN-PH-TH/2007-265\\

\vspace*{2mm}
\end{flushright}
\vspace*{1.5cm}
 
\boldmath
\begin{center}
{\Large{\bf
\prog: a fortran code for $\bratio$ \\

\vspace*{3mm}

in the MSSM with Minimal Flavor Violation}}
\vspace*{5mm}

\end{center}
\unboldmath
\smallskip
\begin{center}
{\large{G.~Degrassi$^{a,\,b}$, P.~Gambino$^c$, and P.~Slavich$^{a,\,d}$}}
\vspace*{8mm} \\
{\sl ${}^a$ CERN, Theory Division, CH-1211 Geneva 23, Switzerland}
\vspace*{2.5mm}\\
{\sl ${}^b$
    Dipartimento di Fisica, Universit\`a di Roma Tre and  INFN, Sezione di
    Roma Tre \\
    Via della Vasca Navale~84, I-00146 Rome, Italy}
\vspace*{2.5mm}\\
{\sl  ${}^c$ Dipartimento di Fisica Teorica, 
Universit\`a di Torino and INFN, Sezione di Torino \\ 
Via P.~Giuria 1,  I-10125 Torino, Italy}  
\vspace*{2.5mm}\\
{\sl ${}^d$  LAPTH, 9, Chemin de Bellevue, F-74941 Annecy-le-Vieux,  France}
\symbolfootnote[0]{{\tt e-mail:}}
\symbolfootnote[0]{{\tt degrassi@fis.uniroma3.it}}
\symbolfootnote[0]{{\tt Paolo.Gambino@to.infn.it}}
\symbolfootnote[0]{{\tt Pietro.Slavich@cern.ch}}

\vspace*{1.5cm}

{\bf Abstract}\vspace*{-.9mm}\\
\end{center}

\noindent
We present the fortran code \theprog\ version \vernum, which computes
the branching ratio for the decay $\Bsg$ in the MSSM with Minimal
Flavor Violation.  The computation takes into account all the
available NLO contributions, including the complete supersymmetric QCD
corrections to the Wilson coefficients of the magnetic and
chromomagnetic operators.

\vspace*{1cm}

{\em \noindent NOTE: this preprint matches the version published in
  Comput.\,Phys.\,Commun.~179 (2008) 759.\\ The latest version of the
  SusyBSG manual can be downloaded from the program's web page.}

\setcounter{page}{0}
\mbox{}
\vfill

\newpage

\section{Introduction}

The radiative $B$ decays play a key role in the program of precision
tests of the Standard Model (SM) and its extensions.  The inclusive
decay $\Bsg$ is particularly well suited to this precision program,
thanks to its low sensitivity to non-perturbative effects. The present
experimental world average for the branching ratio of this decay, with
a 1.6 GeV lower cut on the energy of the photon, is $\bratio_{\rm ex}
= (3.55 \pm 0.26) \times 10^{-4}$ \cite{hfag}. The SM prediction for
the branching ratio with the same cut on the photon energy is
$\bratio_{\rm th} = (3.15 \pm 0.23) \times 10^{-4}$
\cite{supergroup,misiakstein} and includes most of the
next-to-next-to-leading order (NNLO) perturbative QCD contributions as
well as the leading non-perturbative and electroweak effects. Both
experiment and SM prediction have an uncertainty of about 7\%.

New Physics (NP) can in principle induce sizeable contributions to the
decay $\Bsg$, hence the good agreement between the SM prediction and
the experimental result puts severe constraints on the flavor
structure of NP models.  However, the theoretical accuracy of the
predictions for $\bratio$ in extensions of the SM is not at the same
level as in the SM.  A complete next-to-leading order (NLO)
calculation is available only for the Two-Higgs-Doublet Model (THDM)
\cite{CDGG1,2HDM}, the Left-Right symmetric model \cite{bobeth},
and for the Minimal Supersymmetric Standard Model (MSSM) under the
simplifying assumption of Minimal Flavor Violation (MFV), according to
which the quark and squark mass matrices can be simultaneously
diagonalized and the only source of flavor violation is the CKM
matrix.  In this scenario, the diagrams that include gluons and
charginos were computed in refs.~\cite{CDGG2,bobeth}, while the
contributions involving gluinos were first considered in the heavy
gluino limit in ref.~\cite{CDGG2}, and in an effective Lagrangian
approach in refs.~\cite{DGG,carena,DGIS}. After a partial two-loop
calculation \cite{borzu}, the full computation of the two-loop gluino
contributions to $\Bsg$ in the MSSM with MFV was finally presented in
ref.~\cite{DGS}.

Several public computer codes that determine the MSSM mass spectrum
and other SUSY observables (e.g.~{\tt SuSpect} \cite{suspect}, {\tt
SPheno} \cite{spheno}, {\tt micrOMEGAs} \cite{micromegas}, {\tt
FeynHiggs} \cite{feynhiggs}, {\tt NMHDECAY} \cite{nmhdecay}, {\tt
CPsuperH} \cite{cpsuperh} and {\tt SuperIso} \cite{superiso}) contain
calculations of $\bratio$ in various approximations. However, in the
present versions of all these codes the two-loop gluino contributions
to $\Bsg$ are included, if at all, only in the effective Lagrangian
approximation of refs.~\cite{DGG,carena,DGIS}, which is valid in the
limit of heavy superpartners and large $\tan\beta$.

In this paper we present a new fortran code, \theprog, dedicated to
the full NLO calculation of $\bratio$ in the MSSM with MFV. The code
includes the full results of ref.~\cite{DGS} for the two-loop gluino
contributions to the Wilson coefficients of the magnetic and
chromomagnetic operators relevant to the $\Bsg$ decay, and the results
of refs.~\cite{CDGG2,bobeth} for the two-loop gluon contributions.  It
should be recalled that the weak interactions affect the squark and
quark mass matrices in a different way, therefore their simultaneous
diagonalization can be consistently imposed only at a scale $\mumfv$,
which concurs to specify the MFV model. The renormalization group
evolution of the MSSM parameters then leads to a disalignment between
the squark and quark mass matrices at scales different from $\mumfv$.
Thus, optionally, the code allows for the inclusion of (small)
additional contributions to the Wilson coefficients from one-loop
diagrams with gluinos and down-type squarks (as well as charginos and
up-type squarks) that occur when the MFV condition is imposed at a
scale much higher than the weak scale. For the sake of comparison,
\theprog\ can also provide evaluations of $\bratio$ in the SM and in
the THDM, as well as in the MSSM with two-loop gluino contributions
computed in the effective Lagrangian approximation.

In \theprog\ the relation between the Wilson coefficients and the
$\Bsg$ branching ratio is computed at NLO in perturbative QCD, along
the lines of ref.~\cite{GM}, including also the dominant electroweak
and non-perturbative corrections. For new physics that does not induce
effective operators other than those already present in the SM, the
NNLO anomalous dimensions of the effective operators and their matrix
elements are the same as in the SM, suggesting the possibility of a
partial NNLO implementation. A complete NNLO calculation would require
also the NNLO contributions to the matching conditions, which have
been computed only in the SM. However, if it can be argued that these
contributions are as small as in the SM they can safely be
neglected. This has been done for the case of the type II THDM in
ref.~\cite{supergroup}, where the NP contributions to the matching
conditions are computed at NLO and the anomalous dimensions and matrix
elements are computed at NNLO. In the present version of \theprog\ we
follow a similar but simpler route: we take into account the first
NNLO estimate of ref.~\cite{supergroup} by modifying the approach of
ref.~\cite{GM} in a way that approximately reproduces the results of a
partial NNLO implementation. In fact, any attempt at a partial NNLO
implementation has limitations in the MSSM, where higher-order QCD and
electroweak contributions to the matching conditions can be sizeable,
especially at large $\tan\beta$.  The theoretical accuracy of our code
therefore tends to be poorer in the MSSM than in the SM. 

This manual is structured as follows: in section 2 we briefly
summarize the two-loop results implemented in \theprog, focusing on
the information necessary to a correct interpretation of the input
parameters (we refer the readers to ref.~\cite{DGS} for the technical
details on the calculation). In section 3 we describe the structure of
\theprog, focusing in particular on the input and output parameters of
the two main subroutines that make up the program. In section 4 we
briefly detail our default choices for the input parameters and
discuss the theoretical uncertainty.  In the appendices we provide
additional details on the various corrections implemented in \theprog.
In the appendix A we present the one-loop gluino contributions to the
matching conditions for the Wilson coefficients $C_1^{(1)}$ and
$C_2^{(1)}$. In the appendix B we provide formulae for the one-loop
gluino and chargino contributions to the Wilson coefficients in the
presence of flavor mixing in the squark sector.  In the appendix C we
summarize our treatment of the $\tan\beta$-enhanced contributions to
the Wilson coefficients. Finally, in the appendix D we summarize the
NLO computation of the relation between Wilson coefficients and
branching ratio for $\Bsg$.

The latest version of \theprog\ can be downloaded from the Web page

\begin{center}
{\tt http://cern.ch/slavich/susybsg/home.html}
\end{center}

\boldmath
\section{Radiative $B$ decays in the MSSM with MFV}
\unboldmath

In this section we summarize the calculation of the weak-scale
matching conditions for the $\Delta B = 1$ effective Hamiltonian in
the MSSM with Minimal Flavor Violation, as implemented in
\theprog. The NLO relation between the Wilson coefficients at the weak
scale and the branching ratio is summarized for completeness in the
appendix D.

The $\Delta B = 1$ effective Hamiltonian at the matching scale $\mu_0$
(of the order of the weak scale) is given by
\be
\label{effH}
{\cal H}= -\frac{4G_F}{\sqrt{2}}\vckmc_{ts}\vckm_{tb} 
\sum_i C_i(\mu_0) Q_i(\mu_0)\,,
\ee
where $G_F$ is the Fermi constant and $\vckm_{ts},\,\vckm_{tb}$ are elements
of the CKM matrix. The operators relevant to our calculation are
\bea
Q_1 &=& (\bar{s}_L \gamma_\mu T^a c_L)(\bar{c}_L\gamma^\mu T^a b_L)\,,\\
Q_2 &=& (\bar{s}_L \gamma_\mu c_L)(\bar{c}_L\gamma^\mu b_L)\,,\\
Q_3 &=& (\bar{s}_L \gamma_\mu b_L) \sum_q (\bar{q} \gamma_\mu q)\,,\\
Q_4 &=& (\bar{s}_L \gamma_\mu T^a b_L) \sum_q (\bar{q} \gamma_\mu T^a q)\,,\\
Q_5 &=& (\bar{s}_L \gamma_\mu \gamma_\nu \gamma_\rho b_L) 
\sum_q (\bar{q} \gamma^\mu  \gamma^\nu \gamma^\rho q)\,,\\
Q_6 &=& (\bar{s}_L \gamma_\mu \gamma_\nu \gamma_\rho T^a b_L) 
\sum_q (\bar{q} \gamma^\mu  \gamma^\nu \gamma^\rho T^a q)\,,\\
Q_7 &=& \frac{e}{16\pi^2} m_b {\bar s}_L \sigma^{\mu \nu}b_RF_{\mu\nu}\,,\\
Q_8 &=& \frac{g_s}{16\pi^2}\,m_b {\bar s}_L 
\sigma^{\mu \nu}T^ab_RG^a_{\mu\nu}\,.
\eea

When the QCD corrections are considered, the Wilson coefficients of
the operators $Q_i$ can be organized in the following way
\bea
C_i(\mu_0)& = &C^{(0){\,\rm SM}}_i(\mu_0) + 
 C_i^{(0){\,H^\pm}}(\mu_0) + C_i^{(0){\,\susy}}(\mu_0)  \nn\\ 
&&\nn\\
&+& \frac{\as(\mu_0)}{4 \pi} \left[ C^{(1){\,\rm SM}}_i(\mu_0) +
 C_i^{(1){\,H^\pm}}(\mu_0) +  C_i^{(1){\,\susy}}(\mu_0) \right],
\label{wc1}
\eea 
where the various leading order (LO) contributions are classified
according to whether the corresponding diagrams contain only SM
fields, a physical charged Higgs boson and an up-type quark, or a
chargino and an up-type squark.  The expressions for $C_i^{(0){\,\rm
SM}}$ and $C_i^{(0){\,H^\pm}}$ can be found, e.g., in
ref.~\cite{CDGG1}, while those for $C_i^{(0){\,\susy}}$ can be found,
e.g., in eq.~(4) of ref.~\cite{CDGG2}. Note that at LO only the
coefficients of the magnetic and chromo-magnetic operators $Q_7$ and
$Q_8$ receive contributions from non-SM fields. One-loop neutralino-
and gluino-exchange diagrams should be neglected under the MFV
assumption.

The NLO coefficients $C^{(1){\,\rm SM}}_i$ and $C_i^{(1){\,H^\pm}}$
contain the gluonic corrections to the SM and charged Higgs
contributions, respectively, and can be found for instance in
ref.~\cite{CDGG1}. Concerning the NLO supersymmetric contributions
$C_i^{(1){\,\susy}}$, the chargino-gluon contributions can be found in
refs.~\cite{CDGG2,bobeth}, and a full two-loop computation of the
gluino contributions to $C^{(1)\,\susy}_{7,8}$ in the MSSM with
minimal flavor violation was more recently presented in
ref.~\cite{DGS}. Together with the one-loop gluino contributions to
$C^{(1)\,\susy}_{1,2}$ that we present in the appendix A, the results
of ref.~\cite{DGS} provide us with a complete NLO computation of the
supersymmetric QCD contributions to $\bratio$ in the MFV scenario.

The study of NLO contributions in the MFV scenario is complicated by
the fact that the simultaneous diagonalization of squark and quark
mass matrices is not preserved by radiative corrections. As a result,
even when the renormalized mixing matrices for quarks and squarks are
assumed to be flavor-diagonal, flavor-changing counterterms of
electroweak origin have to be taken into account in the two-loop
calculation. The technical issues related to the renormalization of
flavor mixing are discussed in ref.~\cite{DGS}, to which we refer the
interested reader. In the following we summarize the information
necessary to a correct interpretation of the input parameters required
by \theprog.

In the MFV framework the mass matrices for both the up-type and
down-type quarks and squarks are assumed to be simultaneously
flavor-diagonal at some renormalization scale $\mumfv$.  When
computing the matching conditions for the Wilson coefficients we
neglect the masses of the first- and second-generation quarks, as well
as the left-right mixing in the mass matrices of the corresponding
squarks. Thus, the soft SUSY-breaking terms that enter the squark mass
matrices and are relevant to our calculation are: the masses for the
SU(2) squark doublets, $m_{Q_i}$, where $i$ is a generation index; the
masses for the third-generation singlets, $m_T$ and $m_B$; the
trilinear interaction terms for the third-generation squarks, $A_t$
and $A_b$.
We recall that in the so-called super-CKM basis, where the matrices of
Yukawa couplings are diagonal and the squarks are rotated parallel to
the quarks, the $3\times3$ mass matrices for the up-type and down-type
left squarks are related by $(M_U^2)_{LL} = \vckm \,(M_D^2)_{LL}\,
\vckmd$. Therefore, the two mass matrices can be both flavor-diagonal
only if they are flavor-degenerate. This means that the MFV scenario
can be consistently implemented only if we choose a common mass
parameter for the three generations of SU(2) squark doublets,
i.e.~$m_{Q_i} \equiv m_Q$.

Since we are focusing on the QCD corrections at two loops, it is
necessary to specify a renormalization scheme for the input parameters
$m_{Q}, \,m_T$ and $A_t$, which determine the up-type squark masses
and mixing entering $C_{7,8}^{(0){\,\susy}}$ and are subject to $\oas$
radiative corrections. We consider two options: the first is to assume
that they are Lagrangian parameters expressed in a minimal subtraction
scheme such as $\drbar$, at some renormalization scale $\mususy$ of
the order of the superparticle masses. The second is to adopt an
on-shell (OS) definition: $m_{Q}, \,m_T$ and $A_t$ can be interpreted
as the unphysical parameters that enter the tree-level stop mass
matrix obtained by rotating the diagonal matrix of the physical stop
masses by a suitably defined physical mixing angle
$\theta_{\tilde{t}}$. In this case we adopt an OS definition (i.e.~we
use the physical mass) also for the top quark mass that enters the
tree-level stop mass matrix.

The other MSSM parameters relevant to the calculation, for which we
need not specify a renormalization prescription, are: the ratio of
Higgs vacuum expectation values $\tan\beta \equiv v_2/v_1$; the
charged Higgs boson mass $\mhc$; the gluino mass $\mgl$; the SU(2)
gaugino mass parameter $M_2$; the higgsino mass parameter $\mu$. Our
conventions for the signs of the various Lagrangian parameters are as
specified by the {\em SUSY Les Houches Accord} (SLHA) \cite{SLHA,SLHA2},
i.e.: the top and bottom quark masses are $m_t = h_t\,v_2/\sqrt2$ and
$m_b = h_b\,v_1/\sqrt2$, respectively; the left-right mixing terms in
the stop and sbottom mass matrices are $m_t(A_t-\mu\cot\beta)$ and
$m_b (A_b-\mu\tan\beta)$, respectively; the (2,2) entry of the
chargino mass matrix is $\mu$.

\theprog\ has an option to read the parameters of the MSSM Lagrangian
from a SLHA spectrum file. In that case, the parameters are meant to
be expressed in the $\drbar$ renormalization scheme. When a SLHA
spectrum file is available the program reads from it also the
$\drbar$-renormalized top quark mass computed at the scale $\mususy$,
which is necessary for the computation of the running stop masses. In
the absence of a SLHA spectrum file the running top mass is computed
internally by the program, taking into account only the corrections
controlled by the strong gauge coupling.

In a two-loop calculation the interpretation of the MFV requirement
itself depends on the way we renormalize the flavor mixing, i.e.~the
way we fix the counterterms that cancel the divergences of the
antihermitian parts of the quark and squark
wave-function-renormalization (WFR) matrices. In particular, if we
perform a minimal subtraction we are imposing the MFV condition on the
$\drbar$-renormalized parameters of the Lagrangian evaluated at the
scale $\mumfv$. In this scheme $C_{7,8}^{(1){\,\susy}}$ contains
logarithms of the ratio $\msusy/\mumfv$, where $\msusy$ represents the
mass of the superparticles entering the loops. An alternative option
consists in subtracting also the finite part of the antihermitian WFR:
this results in a conventional (and gauge-dependent) on-shell
renormalization scheme \cite{mixing}, in which
$C_{7,8}^{(1){\,\susy}}$ is independent of $\mumfv$.

From the discussion above it is clear that, in models where the MFV
condition is imposed at a scale much larger than the superparticle
masses (such as, e.g., supergravity models where one identifies
$\mumfv$ with the GUT scale), the Wilson coefficients computed in the
minimal subtraction scheme contain very large logarithms of
$\msusy/\mumfv$. In this case, no matter what renormalization scheme
is chosen, the fixed-order calculation does not provide a good
approximation to the correct result. Indeed, in such models the soft
SUSY-breaking mass parameters -- which are flavor-diagonal at the
scale $\mumfv$ -- must be evolved down to $\msusy$ with the
appropriate RGE, thus generating some flavor violation in the squark
mass matrices. When the squark mass matrices are diagonalized, the
resummed logarithms of the ratio $\msusy/\mumfv$ are absorbed in the
couplings of the resulting squark mass eigenstates with the gluinos
(and the charginos and neutralinos). Typically, the effects of the
RGE-induced flavor mixing are relatively small, and we include them
only at leading order\footnote{We follow the approach of
ref.~\cite{bbmr}. For a more complete treatment of the gluino
contributions in scenarios with generic sources of flavor violation
see ref.~\cite{bghw}.}, in the one-loop diagrams with gluinos
and down-type squarks (which would vanish if the MFV condition was
valid at the low scale $\msusy$) and in the one-loop diagrams with
charginos and up-type squarks. The corresponding contributions to the
Wilson coefficients are given explicitly in the appendix B. Once the
logarithmic effects have been taken into account in this way, the
genuine two-loop MFV contributions in $C^{(1)\,\susy}_{7,8}$ can be
computed by setting artificially $\mumfv \sim \msusy$.  In this case,
using either the on-shell scheme or the minimal subtraction scheme to
renormalize the flavor mixing will give basically the same result.

We remark here that the rather complicated task of solving the system
of RGE equations for the soft SUSY-breaking parameters, taking into
account the full flavor structure of the MSSM, is not performed by
\theprog. In order to obtain the squark masses and mixing matrices
that enter the expressions in appendix B, starting from
flavor-universal boundary conditions at the scale $\mumfv$, the users
can run one of the public spectrum calculators that include a full
treatment the flavor structure, such as the latest (still unpublished)
versions of {\tt SPheno} \cite{spheno} or {\tt SoftSusy}
\cite{softsusy}, and pass the results to \theprog\ by means of a SLHA2
\cite{SLHA2} input/output file.

Finally, it is well known that in the MSSM the relation between the
bottom quark mass $m_b$ and the bottom Yukawa coupling $h_b$ is
subject to $\tan\beta$-enhanced threshold corrections \cite{HRS}. If
the bottom Yukawa coupling entering the one-loop part of the Wilson
coefficients is expressed in terms of the SM value of the running
bottom mass, the SUSY threshold corrections induce counterterm
contributions that, although being formally of higher order in the
loop expansion, are enhanced by powers of $\tan\beta$ and may
therefore be sizeable when $\tan\beta$ is large.  As discussed
e.g.~in ref.~\cite{carena1}, such potentially large corrections can be
absorbed in the one-loop results by a suitable redefinition of the
bottom Yukawa coupling. Other $\tan\beta$-enhanced contributions to
the Wilson coefficients appear in the form of corrections to the
Higgs-quark-quark vertices, and have been computed in
refs.~\cite{DGG,carena,DGIS} in an effective Lagrangian approach where
the heavy SUSY particles are integrated out of the theory. Among the
$\tan\beta$-enhanced two-loop contributions, those controlled by the
strong gauge coupling are fully accounted for by the calculation of
ref.~\cite{DGS}. In addition, we include in \theprog\ the
$\tan\beta$-enhanced contributions controlled by the top and bottom
Yukawa couplings, following the approach of ref.~\cite{DGIS}. More
details on the treatment of the $\tan\beta$-enhanced contributions are
given in the appendix C.

\newpage

\section{Structure of the program \theprog}

\theprog\ \vernum\ is structured in two modules, which in principle
could be used independently as stand-alone programs. The first module
is the subroutine {\tt WilsonCoeff}, which computes the weak-scale
matching conditions for the Wilson coefficients of the $\Delta B = 1$
effective Hamiltonian in a physics model to be chosen among the SM,
the Two Higgs Doublet Model and the MSSM. The second module is the
subroutine {\tt getBR}, which computes $\bratio$ at NLO, taking as
input the SM and new-physics contributions to the Wilson coefficients
evaluated by {\tt WilsonCoeff}. The main program {\tt BSGAMMA}, which
can be freely modified according to the users' needs, sets all the
relevant input parameters, calls the two subroutines and prints out
the output.  Within {\tt BSGAMMA} the users are allowed to read the
input parameters required by {\tt WilsonCoeff} from a spectrum file
written in the SLHA format. More specifically, the subroutine {\tt
readSUSY\_SLHA1} reads from a SLHA1 \cite{SLHA} file only the
flavor-conserving parameters required for the calculation of the one-
and two-loop contributions appearing in eq.~(\ref{wc1}). The
subroutine {\tt readSUSY\_SLHA2}, instead, reads from a SLHA2
\cite{SLHA2} file all the parameters of the MSSM Lagrangian, including
the flavor-violating parameters necessary to compute the additional
one-loop contributions discussed in the appendix B. In either case the
spectrum file must be named {\tt SLHA.in}, and it must be located in
the directory where the program is run.

For certain choices of the input parameters it may happen that the
masses of two particles are accidentally very similar to each other,
or that the sum of two masses is very close to a third mass. In some
pathologic cases this leads to numerical instabilities in the output
of \theprog, even if all the formulae for the two-loop gluino
contributions to $\Bsg$ are well behaved when the corresponding limits
are taken analytically.  Since those formulae are very long and depend
in a complicated way on the values of twelve different particle
masses, providing analytical results to cover all the problematic
limits would be highly inefficient in terms of size and speed of the
code. Therefore, we limit ourselves to looking for the occurrence of
accidentally similar masses, and issuing a warning if necessary. We
leave it to the users to check that, in those cases, the result for
$\bratio$ is not unreasonably sensitive to small variations of the
input parameters\footnote{Note that the numerical instabilities can be
greatly reduced by compiling the program in quadruple
precision. Details are provided in the \theprog\ Web page.}. The only
exception is the case in which the mass of the first- or
second-generation up-type squark $m_{\tilde{u}_L}$ is very similar to
the mass of the super-strange $m_{\tilde{s}_L}$. This happens
inevitably if the soft SUSY-breaking parameter $m_Q$, common to both
masses, is much larger than $\mz$. To avoid numerical instabilities we
switch to the analytical results valid in the limit
$m_{\tilde{u}_L}=m_{\tilde{s}_L}$ if the relative difference between
the two masses is less than 1\%.

In the following we describe in detail the input and output parameters
of the two main subroutines that make up \theprog.

\subsection{The subroutine {\tt WilsonCoeff}}
\label{sub1}

The call to the subroutine for the matching conditions to the Wilson 
coefficients reads

\begin{flushleft}
{\tt

~~~~~~~~call WilsonCoeff(imod,scheme,mu0,mususy,mumfv,

~~~~~~~\$~~~~~msq3,mstr,msbr,msql,At,Ab,mHp,mg,M2,mu,tanb,

~~~~~~~\$~~~~~CISM,C7SM,C8SM,CINP,C7NP,C8NP,prob,eqmass)
}
\end{flushleft}

The variables in the first two lines of the {\tt call} command 
are inputs, and are defined as:
\begin{itemize}

\item {\tt integer imod}:~~allows the users to choose the particle
content of the theory and the approximation used in the computation of
the two-loop contribution to the Wilson coefficients.  The values 0--4
correspond to
\begin{itemize}

\item[{0:}] Standard Model fields only. The new-physics contributions 
are set to zero;

\item[{1:}] Two Higgs Doublet Model. The only additional contributions
come from diagrams with a charged Higgs $H^{\pm}$;

\item[{2:}] MSSM with the effective Lagrangian approach. At NLO, only
the $\tan\beta$-enhanced SUSY contributions are included in the Wilson
coefficients as in ref.~\cite{DGIS};

\item[{3:}] MSSM with full NLO QCD contributions. Includes the
two-loop calculation of the gluino contributions presented in
ref.~\cite{DGS}, and the results of ref.~\cite{DGIS} for the
remaining (i.e., non-QCD) $\tan\beta$-enhanced contributions.

\item[{4:}] MSSM with full NLO QCD contributions (as for {\tt imod =
3}), with in addition the contributions of the one-loop diagrams with
gluinos and down-type squarks and chargino and up-type squarks given 
in the appendix B.

\end{itemize} 

\item{\tt logical scheme(2)}:~~contains two logical switches (relevant
only for {\tt imod = 2,3,4}) that allow the users to specify the 
renormalization conditions in the squark sector. In particular:
\begin{itemize}

\item[{1:}] choice of renormalization scheme for the input squark mass 
parameters

({\tt .true.} = OS,~ {\tt .false.} = $\drbar$);

\item[{2:}] choice of renormalization scheme for the MFV condition

({\tt .true.} = OS,~ {\tt .false.} = $\drbar$).

\end{itemize}

If the MSSM Lagrangian parameters are read in from a SLHA file both 
entries of {\tt scheme} should be set to {\tt .false.}

\item{\tt real*8 mu0}:~~renormalization scale $\mu_0$ (of the
order of the weak scale) at which the matching of the Wilson
coefficients is performed.

\item{\tt real*8 mususy}:~~renormalization scale $\mususy$ at
which the input squark mass parameters are given (relevant only for
{\tt scheme(1) = .false.}). 

\item{\tt real*8 mumfv}:~~renormalization scale $\mumfv$ at which the
MFV condition is imposed (relevant only for {\tt scheme(2) =
.false.}).  As explained in section 2, the results of the program are
not reliable if $\mumfv$ is set to be much larger than $\mususy$.

\item{\tt real*8 msq3,mstr,msbr}:~~third-generation squark mass
parameters $m_{Q_3}$, $m_T$ and $m_B$. 

\item{\tt real*8 msql}:~~mass parameter $m_{Q}$ for the first- and
second-generation squark doublets. Note that the MFV condition is only
consistent with $m_{Q} = m_{Q_3}$.

\item{\tt real*8 At,Ab}:~~third-generation Higgs-squark-squark 
interaction terms $A_t$ and $A_b$.

\item{\tt real*8 mHp}:~~mass of the charged Higgs boson $\mhc$.

\item{\tt real*8 mg}:~~gluino mass $\mgl$.

\item{\tt real*8 M2}:~~SU(2) gaugino mass parameter (enters the
chargino masses).

\item{\tt real*8 mu}:~~Higgs-mixing superpotential parameter
$\mu$.

\item{\tt real*8 tanb}:~~ratio of Higgs vacuum expectation values 
$\tan\beta$. 

\end{itemize}
of course, all of the SUSY input parameters are relevant only for {\tt
imod = 2,3,4}, while {\tt mHp} and {\tt tanb} are relevant for {\tt
imod = 1,2,3,4}. The option {\tt imod = 4} requires the presence of a
SLHA2 spectrum file, from which the subroutine {\tt readSUSY\_SLHA2},
to be called before {\tt WilsonCoeff}, reads the SUSY input
parameters. For consistency, when {\tt imod = 4} the entries of {\tt
scheme} should be both {\tt .false.}, and $\mumfv$ should be set close
to $\mususy$.

\vspace{3mm}

The variables in the third line of the  {\tt call} command 
are outputs, and are defined as:
\begin{itemize}

\item{\tt real*8 CISM(6)}:~~vector containing the Standard Model
contributions to $C_i^{(1)}(\mu_0)$ for $1\leq i \leq 6$ (they are
actually different from zero only for $i=1,4$).

\item{\tt real*8 C7SM(2),C8SM(2)}:~~vectors whose two elements are the
Standard Model contributions to $C_{7,8}^{(0)}(\mu_0)$ and
$C_{7,8}^{(1)}(\mu_0)$.

\item{\tt real*8 CINP(6)}:~~vector containing the new-physics
contributions to $C_i^{(1)}(\mu_0)$ for $1\leq i \leq 6$ (in the MSSM
they are different from zero only for $i=1,2,4$).

\item{\tt real*8 C7NP(2),C8NP(2)}:~~vectors whose two elements are the
new-physics contributions to $C_{7,8}^{(0)}(\mu_0)$ and
$C_{7,8}^{(1)}(\mu_0)$.

\item{\tt logical prob}:~~ problem flag, set to {\tt .true.}~if the
relative difference between two (or more) particle masses is less than
1\%. The case $m_{\tilde{u}_L}\approx m_{\tilde{s}_L}$ -- which is
taken care of internally -- is not flagged, nor are some other cases
that never lead to instabilities.

\item{\tt logical eqmass(12)}:~~vector that identifies the masses that
are too close to each other. The ordering is
$(\mt,\,\mw,\,\mgl,\,\mhc,\,m_{\tilde{t}_1},\,m_{\tilde{t}_2}
,\,m_{\tilde{b}_1},\,m_{\tilde{b}_2},\,m_{\tilde u_L},\,m_{\tilde
s_L},\, m_{\chi^+_1},\,m_{\chi^+_2})$, and the entries of {\tt eqmass}
corresponding to the masses that are too close are set to {\tt .true.}

\end{itemize}

In addition to the input variables passed in the call to the
subroutine, the users must set (or read from a SLHA spectrum file) a
number of Standard Model parameters in the common block {\tt SMINPUTS}

\begin{flushleft}
{\tt
~~~~~~~~common/SMINPUTS/mz,mw,mtpole,mbmb,hsm,asmz,azinv
}
\end{flushleft}

The variables in {\tt SMINPUTS} are defined as

\begin{itemize}

\item{\tt real*8 mz,mw}:~~physical masses $\mz$ and $\mw$ for the SM gauge 
bosons.

\item{\tt real*8 mtpole}:~~physical top-quark mass $m_t^{\rm pole}$.

\item{\tt real*8 mbmb}:~~running bottom mass $m_b^{\msbar}(m_b)$ in the
$\msbar$ scheme at the scale $m_b$.

\item{\tt real*8 hsm}:~~mass $m_h$ of the SM-like Higgs boson (used
only for the small electroweak corrections of ref.~\cite{GH}, see
the appendix D).

\item{\tt real*8 asmz,azinv}:~~strong and inverse electromagnetic
couplings $\as(\mz)$ and $1/\alpha(\mz)$ in the $\msbar$ scheme at the
scale $\mz$.

\end{itemize}

\subsection{The subroutine {\tt getBR}}
\label{sub2}

This subroutine computes $\bratio$ at NLO using the results of
refs.~\cite{GM,GH}. For the users' convenience the computation of
the branching ratio will be summarized in the appendix D. The call to
the subroutine reads

\begin{center}
{\tt
call getBR(muw,mut,mub,muc,E0,CINP,C7NP,C8NP,BR)
}
\end{center}

The variable {\tt BR} is the output value for $\bratio$.  All the
other variables in the call are inputs, and are defined as

\begin{itemize}

\item{\tt real*8 muw}:~~weak scale $\muw \sim \mw$ at which the
light quark contributions to the Wilson coefficients are computed.

\item{\tt real*8 mut}:~~weak scale $\mu_t \sim \mt$ at which the top
quark contributions to the Wilson coefficients are computed. The
new-physics contributions (see below) should also be computed by the
subroutine {\tt WilsonCoeff} at a scale $\mu_0 = \mu_t$.

\item{\tt real*8 mub}:~~low scale $\mu_b \sim m_b$ at which the
branching ratio is computed.

\item{\tt real*8 muc}:~~low scale $\mu_c$ at which the charm quark
mass $m_c^{\msbar}(\mu_c)$ entering the SM contributions to the Wilson
coefficients is computed.

\item{\tt real*8 E0}:~~minimum photon energy $E_0$.

\item{\tt real*8 CINP(6)}:~~vector containing the new-physics
contributions to $C_i^{(1)}(\mu_t)$ for $1\leq i \leq 6$ (they are
produced in output by the subroutine {\tt WilsonCoeff}).

\item{\tt real*8 C7NP(2),C8NP(2)}:~~vectors whose two elements are the
new-physics contributions to $C_{7,8}^{(0)}(\mu_t)$ and
$C_{7,8}^{(1)}(\mu_t)$ (they are produced in output by the subroutine
{\tt WilsonCoeff}).

\end{itemize}

In addition to the input variables passed in the call, the subroutine
{\tt getBR} requires the Standard Model input parameters contained in
the common block {\tt SMINPUTS} (see section \ref{sub1}), and a set of
parameters contained in the common block {\tt BRINPUTS}

\begin{center}
{\tt
common/BRINPUTS/a0inv,mcmc,rbs,hlam,ccsl,bsl,lambda,A,rhobar,etabar
}
\end{center}

The variables in {\tt BRINPUTS} are defined as

\begin{itemize}

\item{\tt real*8 a0inv}:~~inverse of the fine-structure constant,
$1/\alpha_{\rm em}$.

\item{\tt real*8 mcmc}:~~running charm mass $m_c^{\msbar}(m_c)$ the $\msbar$
scheme at the scale $m_c$.

\item{\tt real*8 rbs}:~~ratio $m_b/m_s$ entering the gluon
bremsstrahlung contribution.

\item{\tt real*8 hlam}:~~HQET parameter $\lambda_2$ entering the
non-perturbative contribution.

\item{\tt real*8 ccsl}:~~non-perturbative semileptonic phase-space factor $C$.

\item{\tt real*8 bsl}:~~semileptonic branching ratio
${\rm BR}[B \to X_c \,e\, \bar{\nu}]$.

\item{\tt real*8 lambda,A,rhobar,etabar}:~~Wolfenstein parameters 
$\lambda\,,\,A\,,\,\bar{\rho}$ and $\bar{\eta}$ for the CKM matrix.

\end{itemize}

More information on these quantities can be found in the appendix D.
Note that the parameters entering {\tt BRINPUTS} are {\it not} read in
from a SLHA spectrum file (even when one is present) and must be
explicitly set by the users before calling {\tt getBR}.

\subsection{File structure of \theprog}
\vspace{.5cm}

The latest version of \theprog\ can be downloaded from the program's
Web page in the form of a file named {\tt SusyBSG\_vv.tar.gz}, where
{\tt vv} stands for the version number.  When uncompressed and
unpacked, the program consists of the following files:

\begin{itemize}

\item{\tt BSGAMMA.f}:~~contains the main program {\tt BSGAMMA}, to be
modified (or replaced) by the users. It just sets the relevant inputs,
calls the two subroutines {\tt WilsonCoeff} and {\tt getBR} and
prints out the output.

\item{\tt DGStwoloop.f}:~~contains the subroutine {\tt DGSbsgamma},
which computes the two-loop gluino contributions to the Wilson
coefficients, based on the results of ref.~\cite{DGS}. Upon
compilation it includes all the files located in the directory {\tt
source}.

\item{\tt functions.f}:~~contains the definitions of various functions
entering the one-loop and two-loop contributions to the Wilson
coefficients.

\item{\tt getBR.f}:~~contains the subroutine {\tt getBR}, based mostly
on the results of ref.~\cite{GM}, which computes the branching ratio
for the process $\Bsg$ taking as input the new-physics contributions
to the Wilson coefficients.

\item{\tt Makefile}:~by default it uses the compiler {\tt g77} to
produce an executable named \theprog. Further instructions for the
compilation can be found in the program's Web page.

\item{\tt slha2io.f}:~~contains two routines that read the input
parameters (with the exception of those in the common block {\tt
BRINPUTS}) from a spectrum file in the SLHA format
\cite{SLHA,SLHA2}. {\tt readSUSY\_SLHA1} reads only the parameters
necessary to the calculation of the MFV contributions, while {\tt
readSUSY\_SLHA2} reads also the parameters necessary to the
calculation of the one-loop diagrams that involve flavor-violating
gluino-quark-squark or chargino-quark-squark vertices.

\item{\tt SLHA.in}:~~an example of SLHA1 spectrum file that can be
used as input by \theprog. The file was produced with {\tt SoftSusy},
with input parameters corresponding to the so-called SPS1a$^\prime$
\cite{sps1ap} point.

\item{\tt source}:~~the files in this directory contain the explicit
expressions of the two-loop gluino contributions to the Wilson
coefficients, to be included in {\tt DGStwoloop.f} upon compilation.
The files were automatically converted to fortran from {\tt Mathematica}
format using {\tt FormCalc} \cite{formcalc}.

\item{\tt WilsonCoeff.f}:~~contains the subroutine {\tt WilsonCoeff},
which computes the one-loop and two-loop new-physics contributions to
the Wilson coefficients (calling {\tt DGSbsgamma} for the two-loop
gluino contributions).

\end{itemize}

\section{Default values of the SM input parameters}

In a phenomenological study of $\bratio$ in the MSSM with MFV, the SM
input parameters entering the common blocks {\tt SMINPUTS} and {\tt
BRINPUTS} will presumably be fixed once and for all at the beginning
of the calculation.  For the users' convenience we list below the
default values of these parameters in the version \vernum\ of
\theprog. In most cases we use the same values as in
ref.~\cite{misiakstein}, where the users can look for the
corresponding references.

The default values of the SM parameters entering the common block
{\tt SMINPUTS} are:
\begin{center}
\[
\begin{array}{lclcrl}
{\tt mz} &~~=~~& \mz &~~=~~& 91.1876&{\rm GeV}\\
{\tt mw} &~~=~~& \mw &~~=~~& 80.403&{\rm GeV}\\
{\tt mtpole} &~~=~~& m_t^{\rm pole} &~~=~~& 170.9&{\rm GeV}\\
{\tt mb1s} &~~=~~& m_b^{\msbar}(m_b) &~~=~~& 4.2&{\rm GeV}\\
{\tt hsm} &~~=~~& m_h &~~=~~& 115&{\rm GeV}\\
{\tt asmz} &~~=~~& \alpha_s(\mz) &~~=~~& 0.1189&\\
{\tt azinv} &~~=~~& 1/\alpha(\mz) &~~=~~& 127.918&
\end{array}
\]
\end{center}

The default values of the B-physics parameters entering the
common block {\tt BRINPUTS} are:
\begin{center}
\[
\begin{array}{lclcrl}
{\tt a0inv} &~~=~~& 1/\alpha_{\rm em} &~~=~~& 137.036&\\
{\tt mcmc} &~~=~~& m_c^{\msbar}(m_c) &~~=~~& 1.224 &{\rm GeV}\\
{\tt rbs} &~~=~~& m_b/m_s &~~=~~& 50&\\
{\tt hlam} &~~=~~& \lambda_2 &~~=~~& 0.12&{\rm GeV}^2\\
{\tt ccsl} &~~=~~& C &~~=~~& 0.580&\\
{\tt bsl} &~~=~~& {\rm BR}[B \to X_c \,e\, \bar{\nu}]_{\rm exp}
&~~=~~& 0.1061&\\
{\tt lambda} &~~=~~& \lambda &~~=~~& 0.2272&\\
{\tt A} &~~=~~& A &~~=~~& 0.818&\\
{\tt rhobar} &~~=~~& \bar{\rho} &~~=~~& 0.221&\\
{\tt etabar} &~~=~~& \bar{\eta} &~~=~~& 0.340&
\end{array}
\]
\end{center}

The choice of the four independent renormalization scales that appear
in the calculation of the branching ratio and are required as input by
the subroutine {\tt getBR} deserves a separate discussion. We recall
that these scales are: the weak scales $\mut$ and $\muw$, at which we
compute the contributions to the matching conditions for the Wilson
coefficients coming from top and charm quarks, respectively; the low
scale $\mu_b$ at which we compute the matrix elements for the $b\to s
\gamma$ decay; the scale $\mu_c$ at which we express the charm mass
entering the matrix elements.  We adopt the default values
\be
\label{defscales}
\mut = \muw = 2\,\mw\,,~~~\mu_b = 2.5~{\rm GeV}\,,~~~\mu_c = 1~{\rm GeV}.
\ee
Using these values for the scales, and setting $m_t^{pole}=171.4$ GeV,
$m_b^{\msbar}(m_b)=4.15$ GeV and $1/\alpha(\mz) = 128.940$ in order to
reproduce the inputs of ref.~\cite{supergroup}, we obtain a SM
prediction for $\bratio$ of $ 3.15 \times 10^{-4}$, in agreement with
the NNLO result of ref.~\cite{supergroup}. Very good agreement, within
up to 2\%, is also found at various values of $\tan\beta$ and
$m_{H^\pm}$ with the partial NNLO implementation of the THDM from
refs.~\cite{supergroup,misiakprivate}. While we plan to include the
known NNLO corrections in a future version of \theprog, the present
implementation provides an excellent starting point.

It should be noted, however, that $\mu_c$ in eq.~(\ref{defscales}) is
adjusted to a very low value in order to mimic the NNLO contributions
that are not present in our calculation. Therefore, in this case, the
variation of the renormalization scales should not be used to estimate
the intrinsic uncertainty of our calculation. Indeed, the result
of the NLO calculation depends quite sharply on $\mu_c$ around the
value that reproduces the NNLO result.  For example, using $\mu_c =
m_c^{\msbar}(m_c) = 1.224$ GeV, with the other scales fixed as in
eq.~(\ref{defscales}), results in $\bratio = 3.29 \times 10^{-4}$.

Concerning the theoretical uncertainty of our prediction, we recall
that in the SM analysis of refs.~\cite{supergroup,misiakstein} the
error is dominated by a 5\% uncertainty due to unknown $O(\alpha_s
\Lambda_{\rm QCD}/m_b)$ non-perturbative contribution to the matrix
elements. Additional $\sim 4\%$ intrinsic uncertainty stems from the
perturbative part of the calculation and from the estimate of missing
NNLO contributions. The parametric error in the SM is only about 3\%.
All these errors are present in the MSSM calculation as
well. Therefore, we recommend using {\it at least} a 7\% intrinsic
uncertainty, throughout the parameter space, to be added in quadrature
with the parametric uncertainty. Barring the case of subtle
cancellations between various supersymmetric contributions, this
appears to be a realistic guesstimate of our theory uncertainty.

To conclude, we quote the results of \theprog\ \vernum\ in a
characteristic point of the MSSM parameter space, the so called
SPS1a$^\prime$ point \cite{sps1ap}. We used {\tt SoftSusy} to generate
a SLHA1 spectrum file that provides the subroutine {\tt WilsonCoeff}
with the $\drbar$-renormalized SUSY parameters evaluated at a scale
$\mususy$ equal to the geometric average of the stop masses. We set
$\mumfv$ to be equal to $\mususy$, thus enforcing the MFV condition at
the weak scale. We also set the SM input parameters to the default
values listed above (with the exception of $\mw$ and $m_h$, which are
taken from the output of {\tt SoftSusy}), and the renormalization
scales appearing in the calculation of the branching ratio to the
values given in eq.~(\ref{defscales}). For the SM prediction (obtained
with {\tt imod = 0}) we find that $\bratio = 3.16 \times 10^{-4}$. For
the THDM prediction ({\tt imod = 1}) we find $3.91 \times
10^{-4}$. For the MSSM prediction in the effective Lagrangian
approximation ({\tt imod = 2}) we find $2.42 \times 10^{-4}$. Finally,
for the MSSM prediction based on the full NLO QCD calculation ({\tt
imod = 3}) we find $2.60 \times 10^{-4}$.

\vspace*{-3mm}
\section*{Acknowledgements}
\vspace*{-2mm}
We are grateful to B.C.~Allanach, G.~Ganis, U.~Haisch, L.~Hofer,
M.~Misiak, R.~Ruiz and W.~Porod for useful discussions. Special thanks
go to T.~Hahn for useful advice and for providing us with the
diagonalization routines of ref.~\cite{hahndiag}. This work was
supported in part by MIUR under contract 2004021808-009, by an EU
Marie-Curie Research Training Network under contract
MRTN-CT-2006-035505 and by ANR under contract BLAN07-2\_194882.

\newpage

\boldmath
\section*{Appendix A:~ gluino contributions to $C_{1}^{(1)}$ and $C_{2}^{(1)}$}
\begin{figure}[p]
\begin{center}
\mbox{\epsfig{file=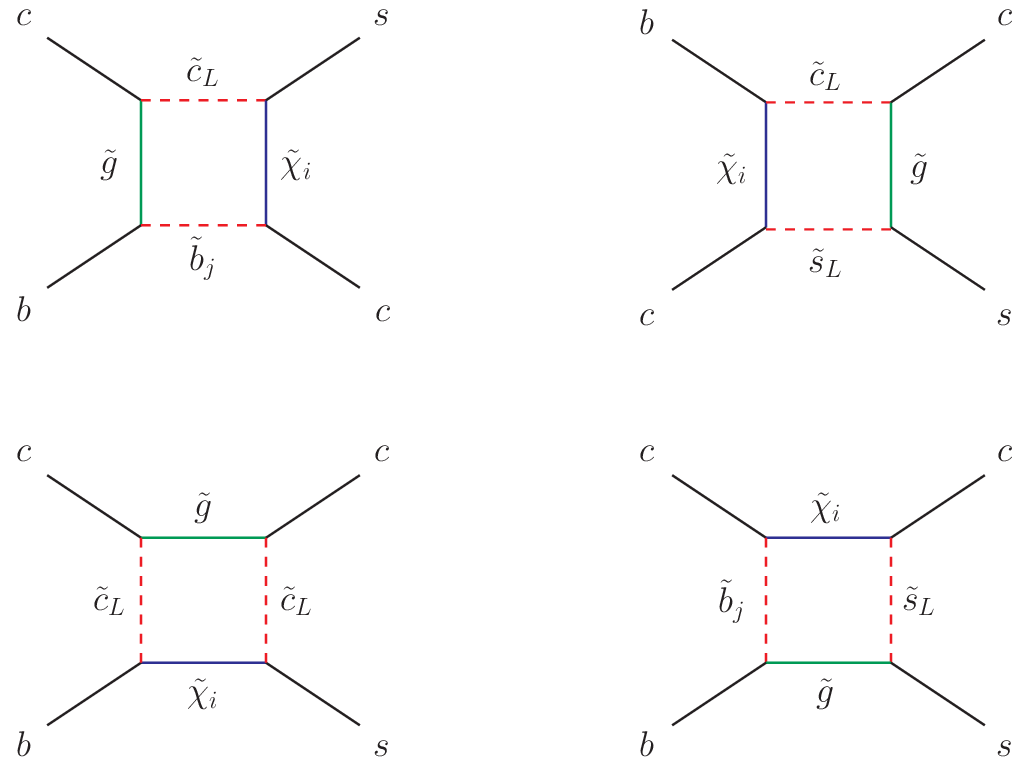,width=14cm}}
\end{center}
\vspace{-2mm}
\caption{\sf Box diagrams contributing to the Wilson coefficient
$C_1^{(1)\,\susy}$.} 
\label{fig1}
\end{figure}

\begin{figure}[p]
\begin{center}
\mbox{\epsfig{file=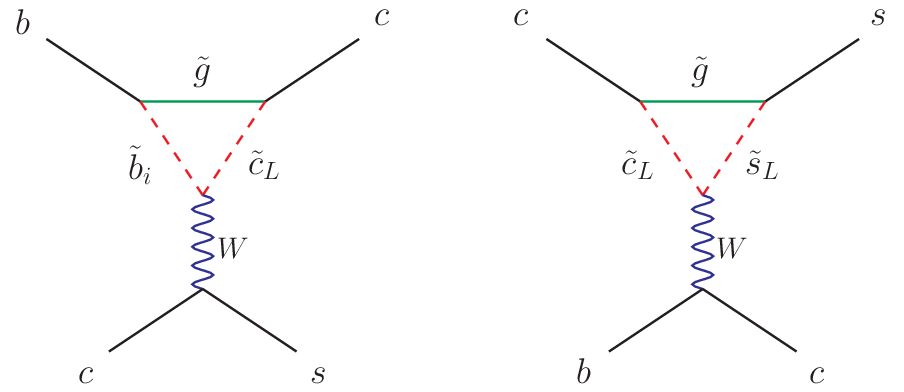,width=13cm}}
\end{center}
\vspace{-2mm}
\caption{\sf Vertex diagrams contributing to the Wilson coefficient
$C_2^{(1)\,\susy}$. Self-energy insertions with a squark-gluino loop on the
external quark legs are also taken into account.}
\label{fig2}
\end{figure}

\unboldmath
\begin{AppendA}
\vspace{.5cm}

A complete NLO computation of the supersymmetric QCD contributions to
$\bratio$ requires the one-loop contributions of ${\cal O}(\as)$ to
the Wilson coefficients of the four-fermion operators $Q_{1-6}$. All
the contributions from Standard Model particles, as well as those
arising from loops involving a charged Higgs boson and a quark or a
chargino and a squark, can be found in the literature
\cite{CDGG1,CDGG2,bobeth}. There are however additional contributions
to the Wilson coefficients $C_{1}^{(1)}$ and $C_{2}^{(1)}$ that
originate from loops involving gluinos, see figs.~\ref{fig1} and
\ref{fig2}. These contributions were neglected in earlier computations
performed under the assumption of heavy gluinos, but they must be
taken into account in a general computation. Assuming MFV, and
neglecting $m_c$ and $m_s$ as well as the left-right mixing in the
charm- and strange-squark sectors, the SUSY contributions to
$C_{1}^{(1)}$ and $C_{2}^{(1)}$ at the matching scale $\mu_0$ read
\bea
C_1^{(1)\,\susy}(\mu_0) & = & 
\frac{\mchi\mw^2}{\mgl^3} \left[
X^b_{j1}\,L^{b\,*}_{ij}\,L^{c\,*}_{i}\,
I_1\left(x_{\chi_i},x_{\tilde{b}_j},x_{\tilde{c}_L}\right)
+L^c_{i}\,L^s_{i}\,
I_1\left(x_{\chi_i},x_{\tilde{s}_L},x_{\tilde{c}_L}\right)\right]\nn\\
&&\nn\\
&&\,+\,\frac{\mw^2}{2\,\mgl^2}\,\left[
\left|L^c_{i}\right|^2\,
I_2\left(x_{\chi_i},x_{\tilde{c}_L},x_{\tilde{c}_L}\right)
+X^b_{j1}\,L^{b\,*}_{ij}\,L^s_{i}\,
I_2\left(x_{\chi_i},x_{\tilde{b}_j},x_{\tilde{s}_L}\right)\right],\nn\\
&&\\
C_2^{(1)\,\susy}(\mu_0) & = & -\frac{C_F}{2}\,\left[
\left|X^b_{i1}\right|^2\, I_3\left(x_{\tilde{b}_i},x_{\tilde{c}_L}\right)
+ I_3\left(x_{\tilde{s}_L},x_{\tilde{c}_L}\right)\right],
\eea
where: summation over the repeated indices is understood; $X^b$ is the
mixing matrix for the sbottoms, defined as
$\;(\tilde{b}_1,\tilde{b}_2)^T \,=\, X^b\,
(\tilde{b}_L,\tilde{b}_R)^T\,$; $C_F = 4/3$ is a color factor; the
quark-squark-chargino couplings are defined as
\be
\label{charcoup}
L^b_{ij} = \frac{m_b}{\mw\cos\beta}\,U_{i2}\,X^b_{j2}
-\sq2\,U_{i1}\,X^b_{j1}\;,~~~~
L^s_{i} = -\sq2\,U_{i1}\;,~~~~
L^c_{i} = -\sq2\,V^*_{i1}\;,
\ee
where $U$ and $V$ are the unitary matrices that diagonalize the
chargino mass matrix according to $U\,{\cal M}_\chi\,V^\dagger = {\rm
diag}\,(m_{\chi_1},m_{\chi_2})$. Finally, $x_P
\equiv m_P^2/\mgl^2\,$ for a generic particle
$P$, and the loop integrals $I_i$ are defined as
\bea
I_1(x_1,x_2,x_3) &=& \frac{x_1\,\ln x_1}{(1-x_1)(x_1-x_2)
(x_1-x_3)}~ + (x_1\leftrightarrow x_2) + (x_1\leftrightarrow x_3)\,,\\
&&\nn\\
I_2(x_1,x_2,x_3) &=& \frac{-x_1^2\,\ln x_1}{(1-x_1)(x_1-x_2)
(x_1-x_3)}~ + (x_1\leftrightarrow x_2) + (x_1\leftrightarrow x_3)\,,\\
&&\nn\\
&&\nn\\
I_3(x_1,x_2) &=& \frac{x_1+x_2-2\,x_1\,x_2}{2\,(x_1-1)(x_2-1)}
+\frac{x_1\,(x_1^2-2\,x_2+x_1\,x_2)\,\ln x_1}{(x_1-1)^2(x_1-x_2)}~ 
+ (x_1\leftrightarrow x_2)\,.
\eea

\end{AppendA}


\section*{Appendix B:~ contributions from squark flavor mixing}
\vspace{.5cm}

\begin{AppendB}

In this appendix we summarize the well-known results for the
gluino--squark and chargino--squark contributions to the Wilson
coefficients in presence of flavor mixing in the squark sector. In the
super-CKM basis, where the squark fields are rotated parallel to their
fermionic partners, the mass eigenstates are linear combinations of
flavor eigenstates, according to
\be
\left(
\begin{array}{c}\tilde u_1\\\tilde u_2\\\tilde u_3\\\tilde u_4\\
\tilde u_5\\\tilde u_6\end{array}\right)~=~
\biggr( \Gamma_{UL}~~\Gamma_{UR}\biggr)
\left(
\begin{array}{c}\tilde u_L\\\tilde c_L\\\tilde t_L\\\tilde u_R\\
\tilde c_R\\\tilde t_R\end{array}\right)~,~~~~~~~
\left(
\begin{array}{c}\tilde d_1\\\tilde d_2\\\tilde d_3\\\tilde d_4\\
\tilde d_5\\\tilde d_6\end{array}\right)~=~
\biggr( \Gamma_{DL}~~\Gamma_{DR}\biggr)
\left(
\begin{array}{c}\tilde d_L\\\tilde s_L\\\tilde b_L\\\tilde d_R\\
\tilde s_R\\\tilde b_R\end{array}\right)~,
\ee
where $\Gamma_{QL}$ and $\Gamma_{QR}$ (with $\scriptstyle Q\,=\,U,D$)
are $6\times3$ mixing matrices. The contributions to the Wilson
coefficients from one-loop diagrams with gluinos and down-type
squarks, $C_{7,8}^{(0){\,\tilde g}}$, have to be added to the first
line in eq.~(\ref{wc1}). They read:
\be
\label{C78g}
C_{7,8}^{(0){\,\tilde g}} = 
\frac{g_s^2\,\mw^2}{g^2\,\mgl^2\,\vckmc_{ts}\vckm_{tb}}\,
\sum_i\left[
\left(\Gamma^*_{DL}\right)_{i2}\left(\Gamma_{DL}\right)_{i3}\,
f_{7,8}\!\left(\frac{m_{\tilde{d}_i}^2}{\mgl^2}\right) -\frac{\mgl}{\mb}\,
\left(\Gamma^*_{DL}\right)_{i2}\left(\Gamma_{DR}\right)_{i3}\,
g_{7,8}\!\left(\frac{m_{\tilde{d}_i}^2}{\mgl^2}\right)\right],
\ee
where $g_s$ and $g$ are the strong and SU(2) gauge couplings,
respectively, and the functions $f_{7,8}$ and $g_{7,8}$ are defined
as:
\be
\label{fun7}
f_7(x) = \frac{2(2+5x-x^2)}{27(x-1)^3}-\frac{4x}{9(x-1)^4}\ln x\,,~~~
g_7(x) = -\frac{4(1+x)}{9(x-1)^2}+\frac{8x}{9(x-1)^3}\ln x\,,
\ee
\be
\label{fun8}
f_8(x) = \frac{(11-40x-19x^2)}{36(x-1)^3}-\frac{x(1-9x)}{6(x-1)^4}\ln x\,,~~~
g_8(x) = -\frac{5-13x}{3(x-1)^2}+\frac{x(1-9x)}{3(x-1)^3}\ln x\,.
\ee

When the information about the mixing in the squark sector is
available we replace the chargino-squark contributions to
$C_{7,8}^{(0){\,\susy}}$ appearing in eq.~(\ref{wc1}) with
\be
\label{C78ch}
C_{7,8}^{(0){\,\susy}} = \sum_{i,j,I,J}
\frac{\vckmc_{Is}\vckm_{Jb}}{\vckmc_{ts}\vckm_{tb}}\,
\left[ - L^{I\,*}_{ij}\,L^J_{ij}\,\frac{\mw^2}{3\,m_{\tilde u_j}^2}\,
F^1_{7,8}\,\left(\frac{m_{\tilde{u}_j}^2}{m_{\chi_i}^2}\right) 
+ L^{I\,*}_{ij}\,R^J_{ij}\,\frac{\mw}{2\,m_{\tilde \chi_i}}\,
F^3_{7,8}\,\left(\frac{m_{\tilde{u}_j}^2}{m_{\chi_i}^2}\right)\right]~,
\ee
where: the indices $\scriptstyle I,\,J$ run over the up-type quark
flavors; the functions $F^1_{7,8}$ and $F^3_{7,8}$ can be read, e.g.,
in eqs.~(2.4) and (3.2)--(3.3) of ref.~\cite{DGG};  the
chargino-quark-squark couplings $L^I_{ij}$ and $R^I_{ij}$ read
\be
L^I_{ij} ~=~ \frac{m_u^I}{\mw\sin\beta}\,V^*_{i2}\,(\Gamma_{UR})_{jI}
- \sqrt{2}\,V^*_{i1}\,(\Gamma_{UL})_{jI}\,,~~~~~~
R^I_{ij} ~=~\frac{1}{\cos\beta}\,U^*_{i2}\,(\Gamma_{UL})_{jI}~,
\ee
where the matrices $U$ and $V$ are defined after
eq.~(\ref{charcoup}). We checked that eqs.~(\ref{C78g}) and
(\ref{C78ch}) agree with the corresponding results in
ref.~\cite{bbmr}.

\end{AppendB}

\newpage

\boldmath
\section*{Appendix C:~ $\tan\beta$-enhanced contributions}
\unboldmath
\begin{AppendC}
\vspace{.5cm}

In this appendix we describe in detail the treatment of the
$\tan\beta$-enhanced contributions to the Wilson coefficients relevant
to $\bsg$. The contributions arising from the threshold corrections to
the relation between the bottom mass and the bottom Yukawa coupling
\cite{HRS} can be absorbed in the one-loop part of the Wilson
coefficients by a suitable redefinition of the Yukawa coupling
$h_b$. Indicating by $-\epsilon_b$ the $\tan\beta$-enhanced part of
the SUSY contribution to the bottom quark self-energy and by
$-\delta_b$ the rest of the SUSY contribution, so that the relation
between the running bottom masses in the SM and in the MSSM is
\be
m_b^{\rm SM} = m_b^{\rm MSSM}(1  + \epsilon_b \tan\beta + \delta_b)\,,
\ee
we multiply the contributions of the one-loop diagrams that involve a
bottom Yukawa coupling by a factor $\kappa$, defined as
\be
\label{defK}
\kappa = \frac{1-\delta_b}{1  + \epsilon_b \tan\beta}\,.
\ee
This amounts to defining $h_b$ in terms of the MSSM running bottom
mass in the one-loop contributions to the Wilson coefficients, with
the result that the counterterm contributions at higher orders do not
contain terms enhanced by powers of $\tan\beta$.  For consistency we
multiply by $\kappa$ also the contributions of the two-loop diagrams
that involve a bottom Yukawa coupling, but we neglect there the small
effect of $\delta_b$.
The $\oas$ contributions to $\epsilon_b$ and $\delta_b$ read
\bea 
\label{epsbs}
\epsilon_b^{(s)} &=& -\frac{\as(\mu_0)}{3\pi}\,
\frac{2\,\mu}{\mgl}\,H_2\left(\frac{\msbu}{\mgl^2},
\frac{\msbd}{\mgl^2}\right)\\
&&\nn\\ \delta_b^{(s)} &=& \frac{\as(\mu_0)}{3\pi}\,
\biggr\{\frac{2\,A_b}{\mgl}\,
H_2\left(\frac{\msbu}{\mgl^2},\frac{\msbd}{\mgl^2}\right)
-\biggr[B_1(0,\mgl^2,\msbu,\mu_0^2)+B_1(0,\mgl^2,\msbd,\mu_0^2)\biggr]
\,\biggr\}\,,
\eea
where $H_2(x,y)$ is defined in eq.~(2.8) of ref.~\cite{DGG}, and 
$B_1(p^2,m_1^2,m_2^2,\mu^2)$ is a Passarino-Veltman function defined as in 
ref.~\cite{PBMZ}.

In addition to the $\tan\beta$-enhanced contributions that arise from
the threshold corrections to the bottom mass, there are
$\tan\beta$-enhanced contributions arising from corrections to the
Higgs-quark-quark vertices \cite{DGG,carena,DGIS}. We stress that all
the contributions of this kind controlled by the strong gauge coupling
are fully accounted for by the explicit two-loop calculation of
ref.~\cite{DGS}.

The $\tan\beta$-enhanced contributions to the Wilson coefficients
controlled by the top and bottom Yukawa couplings are implemented in
\theprog\ using the effective Lagrangian approach of
refs.~\cite{DGG,carena,DGIS}. They include a contribution of $\oat$ to
the bottom quark self-energy
\be
\label{epsbt}
\epsilon_b^{(t)} =
-\frac{\at(\mu_0)}{4\pi}\,
\frac{A_t}{\mu}\,H_2\left(\frac{m_Q^2}{\mu^2},\frac{m_T^2}{\mu^2}\right)\,,
\ee
such that we must use $\epsilon_b = \epsilon_b^{(s)} +
\epsilon_b^{(t)}$ when computing the factor $\kappa$ in eq.~(\ref{defK}).
In addition, there is a contribution to the Wilson coefficients
arising from a correction of $\oat$ to the effective $G^+$--$t$--$b$
vertex:
\be
\label{pino}
\delta C_{7,8} = 
- \frac{\epsilon_b^{(t)}\,\tan\beta}{1+\epsilon_b\,\tan\beta}\,
F^{(2)}_{7,8}\left({m_t^2}/{\mw^2}\right)\,,
\ee
where the loop functions $F^{(2)}_{7,8}(x)$ are defined e.g.~in
eq.~(2.4) of ref.~\cite{DGG}. In the limit of heavy superparticles the
contribution in eq.~(\ref{pino}) cancels an analogous term originating
from the redefinition of the bottom Yukawa coupling, thus ensuring the
decoupling of new-physics effects from the SM contribution.

There are also contributions to the Wilson coefficients involving the
bottom Yukawa coupling and higher powers of $\tan\beta$:
\bea
\label{gino}
\delta C_{7,8} &=& 
- \frac{\epsilon_b^{(t)}\,\epsilon_t^{(b)}\,\tan^2\beta}
{(1+\epsilon_b\,\tan\beta)(1+\epsilon_b^{(s)}\,\tan\beta)}\,
F^{(2)}_{7,8}\left({m_t^2}/{\mhc^2}\right)\nn\\
&& - \frac{\epsilon_b^{(t)}\,\tan^3\beta}
{(1+\epsilon_b\,\tan\beta)^2(1+\epsilon_b^{(s)}\,\tan\beta)}\,
\frac{m_b^2}{36\,\mhc^2}\,,
\eea
where the $\oab$ vertex correction $\epsilon_t^{(b)}$ is defined as
\be
\label{epstb}
\epsilon_t^{(b)} =
\frac{\ab(\mu_0)}{4\pi}\,
\frac{A_b}{\mu}\,H_2\left(\frac{m_Q^2}{\mu^2},\frac{m_B^2}{\mu^2}\right)\,.
\ee
Being suppressed by an additional loop factor and by $m_b^2/\mhc^2$,
respectively, the two terms in eq.~(\ref{gino}) tend to be numerically
small, but, as stressed in ref.~\cite{DGIS}, they can become relevant
in special cases where the leading $\oas$ and $\oat$ effects cancel
each other.

When implementing in \theprog\ the $\tan\beta$-enhanced contributions
controlled by the top and bottom Yukawa coupling,
eqs.~(\ref{epsbt})--(\ref{epstb}), we have neglected the mixing between
superpartners, approximating the squark masses with the corresponding
soft SUSY-breaking terms and the masses of the higgsino components of
charginos and neutralinos with the superpotential parameter
$\mu$. Indeed, the effective Lagrangian approach used to derive these
results relies on the assumption that the superpartners are much
heavier than the weak scale and can be integrated out of the theory,
leaving behind only non-decoupling corrections to the
Higgs-quark-quark vertices. In this case the effect of the mixing,
which is due to the breaking of electroweak symmetry, can reasonably
be expected to be small, and anyway is of the same order of magnitude
as other effects that are neglected under the effective Lagrangian
approximation\footnote{Formulae for the $\tan\beta$-enhanced
contributions to the Wilson coefficients that take into account the
effect of superpartner mixing have been presented in
refs.~\cite{DGG,micromegas,buras2,gomez}.}.

\end{AppendC}

\newpage

\boldmath
\section*{Appendix D:~ NLO determination of $\bratio$}
\unboldmath
\begin{AppendD}
\vspace{.5cm}

In this appendix we summarize the NLO computation of the relation
between the branching ratio for $\Bsg$ and the Wilson coefficients
computed at the electroweak scale, as implemented in \theprog. We rely
mainly on the results of ref.~\cite{GM}, but we also take into account
the matrix elements for the four-fermion operators computed in
ref.~\cite{burasetal} (resulting in an updated table of ``magic
numbers'') and the electroweak matching conditions computed in
ref.~\cite{GH}.

The $\Bsg$ branching ratio with a photon energy cutoff $E_0$ in the
$B$-meson rest frame can be related to the experimentally measured
rate ${\rm BR}[B \to X_c \,e\, \bar{\nu}]_{\rm exp}$ as
\be
\label{defBR} 
\frac{\bratio_{E_{\gamma} > E_0}}
{{\rm BR}[B \to X_c \,e\, \bar{\nu}]_{\rm exp}}
~=~  \left| \frac{ \vckmc_{ts} \vckm_{tb}}{\vckm_{cb}} \right|^2 
\f{6 \,\alpha_{\rm em}}{\pi\;C} 
\left[\, \left|\, K \right|^2 + B(E_0) + N(E_0) \right]\,,
\ee
where $C$ is the ``non-perturbative semileptonic phase-space factor''
defined in the appendix C of ref.~\cite{GM}. The quantities $B(E_0)$
and $N(E_0)$ represent the gluon-bremsstrahlung and non-perturbative
contributions, respectively, and will be discussed below. The factor
$K$ contains the contributions to the $b\to s \gamma$ amplitude and is
dominated by the effective Wilson coefficient for the magnetic
operator at the low scale. It can be written as
\be
\label{c7nlo}
K = K_c + 
r(\mu_t)\, \left[ K_t + K_{\rm NP}\right] + \varepsilon_{\rm ew}\,. 
\ee
Following ref.~\cite{GM} we separate $K$ into the light quark
contribution $K_c$, the top-quark contribution $K_t$, the new-physics
contribution $K_{\rm NP}$ and the electroweak correction
$\varepsilon_{\rm ew}$. In the top and NP contributions we keep the
bottom Yukawa coupling renormalized at the matching scale $\mu_t$ by
introducing the quantity $r(\mu_t)$, defined as
\be 
\label{defr}
r(\mu_t) \equiv \f{m_b^{\overline{\rm MS}}(\mu_t)}{m_b^{1S}}
~=~\f{m_b^{\overline{\rm MS}}(\mu_t)}{m_b^{\overline{\rm MS}}(\mu_b)}
~\times~ \left[1 - \frac{\as(m_b^{1S})}{4\pi}\left(
\frac{16}{3} + 8 \ln \frac{m_b^{1S}}{\mu_b}\right)
+ \frac29 \, \as(m_b^{1S})^2\right]\,,
\ee
where the ``1S mass'' $m_b^{1S}$ is defined as half of the
perturbative contribution to the $\Upsilon$-mass. We extract the value
of $m_b^{1S}$ from the input value of $m_b^{\overline{\rm MS}}(m_b)$
using eq.~(168) of ref.~\cite{hoang}. Unlike ref.~\cite{GM}, we
evaluate $r(\mu_t)$ strictly at NLO. Indeed, in eq.~(\ref{defr}) the
ratio $r(\mu_t)$ is expressed as the product of a term that accounts
for the evolution of the running bottom mass from $\mu_t$ to $\mu_b$
and a term (in the square brackets) that accounts for the shift
between $m_b^{\msbar}(\mu_b)$ and $m_b^{1S}$.  The NLO relation
between the values of a running quark mass evaluated at two different
renormalization scales $\mu$ and $\mu_0$ reads:
\be
\label{runquark}
m_q^{\msbar} (\mu )=m_q^{\msbar} (\mu_0) 
\left[ \frac{\as (\mu)}{\as (\mu_0)}
\right]^{\frac{\gamma_0}{2\beta_0}}
\left[ 1+ \frac{\as (\mu_0)}{4\pi}\frac{\gamma_0}{2\beta_0}\left(
\frac{\gamma_1}{\gamma_0}-\frac{\beta_1}{\beta_0}\right)
\left( \frac{\as (\mu )}{\as (\mu_0)}-1\right) \right]\,,
\ee
where $\beta_0 = 11 - \frac{2}{3}\,n_f\,,~\beta_1 = 102 -
\frac{38}{3}\,n_f\,,~ \gamma_0 =8$ and $\gamma_1
=\frac{404}{3}-\frac{40}{9}\,n_f$ ($n_f$ being the number of active
quark flavors). The strong gauge coupling $\as(\mu)$ is extracted at
NLO from $\as(\mz)$ according to
\be
\label{astrong}
\as(\mu)=\frac{\as (\mz)}{v}\left( 1-\frac{\beta_1}{\beta_0}
\frac{\as (\mz)}{4\pi}\frac{\ln v}{v}\right),~~~~~~
v=1+\beta_0\frac{\as (\mz)}{2\pi}\ln \frac{\mu}{\mz}\,.
\ee

The light-quark contribution to $K$ reads
\bea
K_c &=& -\frac{23}{36}\,\eta_\smallw^{\f{16}{23}} 
-\f{8}{9}\,\left( \eta_\smallw^{\f{14}{23}} 
- \eta_\smallw^{\f{16}{23}}\right) \,
+ \sum_{k=1}^8 h_k \,\eta_\smallw^{a_k}\nn\\
&&\nn\\
&+& \f{\as(\mu_b)}{4\pi}\, \sum_{k=1}^8 
\eta_\smallw^{a_k} \left[\frac{46}{3} \,
a_k \,d_k \,\left( \ln \f{m_b^{1S}}{\mu_b} 
+ \eta _\smallw\ln \f{\muw}{\mw} \right)
+ \tilde{d}_k + \tilde{d}^{\,\eta}_k \, \eta_\smallw
+ \tilde{d}^{\,a}_k\, a(z) 
+ \tilde{d}^{\,b}_k\, b(z) \right] \nn\\
&&\nn\\
\label{kappac}
&+& \f{\vckmc_{us} \vckm_{ub}}{\vckmc_{ts} \vckm_{tb}} 
\, \f{\as(\mu_b)}{4\pi} \,
\left( \eta_\smallw^{a_3} + \eta_\smallw^{a_4} \right)\,  [ a(z) + b(z) ]\,,
\eea
where: $\muw$ is the matching scale (not necessarily equal to
$\mu_t$); $\eta_\smallw = \as(\muw)/\as(\mu_b)$; the ``magic numbers''
$h_k$ are given in eq.~(2.3) of ref.~\cite{GM}; the ``magic numbers''
$a_k,d_k,\tilde{d}_k,\tilde{d}^{\,a}_k$ and $\tilde{d}^{\,b}_k$ are
given in table 2 of ref.~\cite{burasetal}; the functions $a(z)$ and
$b(z)$ are given in eqs.~(D.1) and (D.2) of ref.~\cite{GM}. The
variable $z$ is defined as $(m_c^{\msbar}(\mu_c)/m_b^{1S})^2$, where
$\mu_c$ is an arbitrary renormalization scale.

After factoring out $r(\mu_t)$, the top-quark contribution to $K$
reads
\bea
K_t &=& \left[ 1 -\f{2}{9}\, \as(m_b^{1S})^2 \right]\, \left[
\eta_t^{\f{4}{23}} \,C_{7~~ {\rm top}}^{(0)\,{\rm SM}}(\mu_t) 
+\f{8}{3}\,\left( \eta_t^{\f{2}{23}} - \eta_t^{\f{4}{23}}\right)\, 
C_{8~~ {\rm top}}^{(0)\,{\rm SM}}(\mu_t)\right]\nn\\
&&\nn\\
&+& \f{\as(\mu_b)}{4\pi} 
\left\{\;\sum_{k=1}^8 e_k \,\eta_t^{\left(a_k+\f{11}{23}\right)}\,
C_{4~~ {\rm top}}^{(1)\,{\rm SM}}(\mu_t)\right. \nn\\ 
&&\nn\\
&& \left. \!\!+ \eta_t^{\f{4}{23}} \left[
~~\eta_t\, C_{7~~{\rm top}}^{(1){\,\rm SM}}(\mu_t)
- 2\, \left( \f{12523}{3174} -\f{7411}{4761}\,\eta_t -\f{2}{9} \pi^2
-\f{4}{3}\,\ln \f{m_b^{1S}}{\mu_b}\right) 
\,C_{7~~{\rm top}}^{(0){\,\rm SM}}(\mu_t)\right. \right. \nn\\ 
&&\nn\\
&& ~~~~~~\,
\left. \left. -\f{8}{3}\,\eta_t\, C_{8~~{\rm top}}^{(1){\,\rm SM}}(\mu_t)
-2\, \left( -\f{50092}{4761} +\f{1110842}{357075} \eta_t +\f{16}{27} \pi^2
+\f{32}{9}\,\ln \f{m_b^{1S}}{\mu_b} \right)
\,C_{8~~{\rm top}}^{(0){\,\rm SM}}(\mu_t)\right]\right. \nn\\ 
&&\nn\\
&& \left. \!\!+ \eta_t^{\f{2}{23}} \left[ 
~~\f{8}{3}\,\eta_t \,C_{8~~{\rm top}}^{(1){\,\rm SM}}(\mu_t)
-2 \left( \f{2745458}{357075} -\f{38890}{14283}\,\eta_t -\f{4}{9} \pi^2
-\f{16}{9}\,\ln \f{m_b^{1S}}{\mu_b}\right)\,
C_{8~~{\rm top}}^{(0){\,\rm SM}}(\mu_t)\right] \right\}~,\nn\\ 
\label{ktop}
\eea
where: $\eta_t = \as(\mu_t)/\as(\mu_b)$; the ``magic numbers'' $e_k$
are given in table 2 of ref.~\cite{burasetal}; the top quark
contributions to the matching conditions for the Wilson coefficients
at the scale $\mu_t$ can be expressed in terms of the functions
$E_0^t\,,A_{0,1}^t$ and $F_{0,1}^t$ appearing in eqs.~(6), (10)--(12)
and (17) of ref.~\cite{BMU} as
\be
\label{funbob}
C_{4~~{\rm top}}^{(1){\,\rm SM}}(\mu_t) = E_0^t(x)\,,~~~~
C_{7~~{\rm top}}^{(0,1){\,\rm SM}}(\mu_t) = -\hlf\,A_{0,1}^t(x)\,,~~~~
C_{8~~{\rm top}}^{(0,1){\,\rm SM}}(\mu_t) = -\hlf\,F_{0,1}^t(x)\,,
\ee
where $x = (m_t^{\msbar}(\mu_t)/\mw)^2$. To compute $x$, the
$\msbar$-renormalized running top mass is extracted from the pole mass
$m_t^{\rm pole}$ according to
\be
m_t^{\msbar} (m_t^{\rm pole}) = 
m_t^{\rm pole}\;\left[1 + \frac{4}{3}\,\frac{\as(m_t^{\rm pole})}{\pi}
+ 10.9\,\frac{\as^2(m_t^{\rm pole})}{\pi^2}\right]^{-1},
\ee
and is then evolved to the renormalization scale $\mu_t$ according to
eq.~(\ref{runquark}).

The new-physics contribution to $K$ can be obtained by simply
replacing the top contributions in eq.~(\ref{ktop}) with the
corresponding NP contributions. In the MSSM one has to add also the
squark-gluino contributions to the Wilson coefficients $C_1$ and
$C_2$, presented in the appendix A.  In summary,
\be
\label{kappanp}
K_{\rm NP} = 
K_t ~~\left(C^{(i) \,{\rm SM}}_{n~{\rm top}} 
\rightarrow C^{(i)\,{\rm NP}}_n \right)
~~+~~ \f{\as(\mu_b)}{4\pi}\,\sum_{k=1}^8 \, 
\left[f_k\,C_1^{(1)\,{\rm NP}}(\mu_t)
+h_k \,\,C_2^{(1)\,{\rm NP}}(\mu_t)\right] 
\,\eta_t^{\left(a_k+\f{11}{23}\right)}
\ee
where the ``magic numbers'' $h_k$ are the same as in $K_c\,$, and the 
$f_k$ are:
\be
f_k = (0.5784,~-0.3921,~-0.1429,\,\,~~~~0.0476,~-0.1275,~~~~\,0.0317,
~~~~\,0.0078,~-0.0031).
\ee

In the determination of the electroweak contribution $\varepsilon_{\rm
ew}$ that appears in eq.~(\ref{c7nlo}) we adopt the results of
ref.~\cite{GH} [see in particular eq.~(3.10) there], including the
available NP contributions where appropriate.

The gluon-bremsstrahlung contribution to the branching ratio is
\be
B(E_0) = \f{\as(\mu_b)}{\pi} 
\sum_{\begin{array}{c} \ 
\\[-7mm] {\scriptscriptstyle i,j = 1, ..., 8} \\
[-2mm] {\scriptscriptstyle i \leq j} \end{array}}
C_i^{(0)\,\rm eff}(\mu_b) \; C_j^{(0)\,\rm eff}(\mu_b) \; \phi_{ij} (\delta)\,,
\ee
where: $\delta \equiv 1 - 2 E_0/\mb^{1S}$; the expression for the
functions $\phi_{ij}(\delta)$ are given in eqs.~(E.2)--(E.8) of
ref.~\cite{GM}; the ratio $m_b/m_s$ appearing in the expression of
$\phi_{88}(\delta)$ is taken as input. The contributions corresponding
to $3 \leq i \leq 6$ or $3 \leq j \leq 6$ are negligible and we set
them to zero. We take the SM contributions to the LO effective Wilson
coefficients $C_i^{(0)\,\rm eff}(\mu_b)$ from eq.~(E.9) of
ref.~\cite{GM}, and include the NP contributions where appropriate.

Finally, the non-perturbative contribution to the branching ratio can
be approximated as (see \cite{GM} for the complete list of references)
\be
N(E_0) ~\simeq~ -\frac{1}{18}\,C_7^{(0)\,\rm eff}(\mu_b)\,
\left[\,\eta_\smallw^{\frac{6}{23}} +  \eta_\smallw^{-\frac{12}{23}}\,\right]\,
\frac{\lambda_2}{m_c^2}\,,
\ee
Where the parameter $\lambda_2$ is taken as input, and for the charm
quark mass we use $m_c^{\msbar}(m_c)$.

\end{AppendD}

\newpage


\noindent{\bf PROGRAM SUMMARY}\\
\begin{small}
{\em Manuscript Title:}~ \prog: a fortran code for $\bratio$ 
in the MSSM with Minimal Flavor Violation \\
{\em Authors:}~ G.~Degrassi, P.~Gambino, P.~Slavich\\
{\em Program Title:}~ \prog\\
{\em Journal Reference:}                                      \\
{\em Catalogue identifier:}                                   \\
{\em Licensing provisions:}~ None\\
{\em Programming language:}~Fortran\\
{\em Operating system:}~Linux\\
{\em Keywords:}~Supersymmetry, B physics, rare decays   \\
{\em PACS:}~12.60.Jv, 13.20.He\\
{\em Classification:}~11.6 Phenomenological and Empirical Models and Theories\\
%
 {\em Nature of problem:}\\
Predicting the branching ratio for the decay $\Bsg$ in the
MSSM with Minimal Flavor Violation.   \\
{\em Solution method:}\\ 
We take into account all the available NLO contributions, including
the two-loop gluino contributions to the Wilson coefficients computed
in ref.~[1]. The relation between the Wilson coefficients and the
branching ratio for $\Bsg$ is computed along the lines of
ref.~[2]. The SUSY input parameters can be read from a spectrum file
in the SLHA format.  \\
{\em Restrictions:}\\
The results apply only to the case of Minimal Flavor Violation.\\
{\em Unusual features:}\\
Numerical instabilities may arise for special combinations of the
input parameters. They can usually be avoided by compiling the code in
quadruple precision.\\
{\em Running time:}\\
Approximately 10 ms in double precision, 0.5 s in quadruple precision
(on a Pentium D 3.40-GHz).\\
{\em References:}

\begin{refnummer}
\item          
  G.~Degrassi, P.~Gambino and P.~Slavich,
  Phys.\ Lett.\ B {\bf 635} (2006) 335
  [arXiv:hep-ph/0601135].
\item 
  P.~Gambino and M.~Misiak,
  Nucl.\ Phys.\ B {\bf 611} (2001) 338
  [hep-ph/0104034].
\end{refnummer}

\end{small}

\end{document}